\documentclass[prl,twocolumn,superscriptaddress,showpacs,%
preprintnumbers,amsmath,amssymb,floatfix]{revtex4}
\usepackage{times}
\usepackage{graphicx}
\usepackage{dcolumn}
\usepackage{bm}
\usepackage{subfigure}

\begin{document}
\title{Earthquake Forecast via Neutrino Tomography}

\author{Bin Wang}
\author{Ya-Zheng Chen}
\author{Xue-Qian Li}
\affiliation{Department of Physics, Nankai University, Tianjin,
300071, China}

\begin{abstract}
We discuss the possibility of forecasting earthquakes by means of
(anti)neutrino tomography. Antineutrinos emitted from reactors are
used as a probe. As the antineutrinos traverse through a region
prone to earthquakes, observable variations in the matter effect on
the antineutrino oscillation would provide a tomography of the
vicinity of the region. In this preliminary work, we adopt a
simplified model for the geometrical profile and matter density in a
fault zone. We calculate the survival probability of electron
antineutrinos for cases without and with an anomalous accumulation
of electrons which can be considered as a clear signal of the coming
earthquake, at the geological region with a fault zone, and find that
the variation may reach as much as 3\% for $\bar \nu_e$ emitted
from a reactor. The case for a $\nu_e$ beam from a neutrino factory
is also investigated, and it is noted that, because of the typically
high energy associated with such neutrinos, the oscillation length
is too large and the resultant variation is not practically
observable. Our conclusion is that with the present reactor
facilities and detection techniques, it is still a difficult task to
make an earthquake forecast using such a scheme, though it seems to
be possible from a theoretical point of view while ignoring some
uncertainties. However, with the development of the geology,
especially the knowledge about the fault zone, and with the
improvement of the detection techniques, etc., there is hope that
a medium-term earthquake forecast would be feasible.
\end{abstract}

 \maketitle


{\em Introduction.---}Earthquakes and tsunamis are natural catastrophes.
They occur on our
globe so frequently and cost thousands of lives and immense loss of
property.
We lack an efficient way to forecast earthquakes at present. We believe
that our knowledge of modern physics and sophisticated detection techniques
and facilities may help. It has been proposed that we could
explore oil storage with neutrino beams
\cite{{oil-Glashow},{oil-Smirnov},{oil-Winter}}, which also find
applications in various other fields
\cite{{B-L.Young},{various-applications}}. These creative ideas
motivate us to investigate the possibility of forecasting
earthquakes and tsunamis in terms of matter effects on
(anti)neutrino oscillations.


Various neutrino experiments are, or will soon be, in operation. With great
effort, most neutrino parameters have been determined with certain
accuracy. However, the complex phase $\delta_{\rm CP}$ and the sign
of $\Delta {m}^2_{31}$ are still unknown. Moreover, so far we only
have the upper bound of $\theta_{13}$, which is manifestly smaller
compared with the other two mixing angles. Precise measurement of
$\theta_{13}$ will be valuable for searches for CP violation in the
lepton sector. Two reactor neutrino experiments, Daya Bay
\cite{Daya-Bay} in China and Double Chooz \cite{Double-Chooz} in
France, aiming to measure $\theta_{13}$ are expected to reach very
high precision at the percent level.

Matter effects on the neutrino oscillation are crucial in some
cases, for example, to explain the solar neutrino flux deficit.
Wolfenstein \cite{Wolfenstein} formulated the matter effect on
neutrino oscillation and then Mikheev and Smirnov \cite{MS} applied
and developed the theory to successfully solve the solar neutrino
problem. The newest experimental data confirm the MSW model.

As a neutrino traverses through matter, its oscillation length is no
longer the same as in vacuum. Electron neutrinos acquire an extra
potential $V(x)=\sqrt{2}G_Fn_e(x)$
in the Hamiltonian due to their interaction with electrons, which
exist in ordinary matter, where $G_F$ is the Fermi coupling constant
and $n_e$ is the number density of electrons in matter. In a
simplified version with only two generations,  the expression of the
survival probability in uniform matter is similar to the vacuum
case, but the vacuum mixing angle $\theta_0$ and mass-squared
difference $\Delta m^2 =|m_2^2-m_1^2|$ should be replaced by the
corresponding quantities in matter. That is
\begin{eqnarray}
\ P_{\nu_e\rightarrow\nu_e}=1-\sin^2{2\theta_M}\sin^2\left({\Delta
m_M^2\over4E_{\nu}}L\right), \label{eq:pnue->nue}
\end{eqnarray}
where $\theta_M$ and $\Delta m^2_M$ are the mixing angle and
mass-squared difference in matter, respectively. $E_{\nu}$ and $L$
are, respectively, the energy of the neutrino and the length of the
baseline. For $\theta_M$ we have
\begin{eqnarray}
\tan{2\theta_M}=\tan{2\theta_0}\left({1-{2\sqrt{2}E_\nu G_Fn_
e\over\Delta m^2\cos2\theta_0}}\right)^{-1}.\label{eq:tan2thetaM}
\end{eqnarray}
For antineutrinos, the minus sign in Eq.~(\ref{eq:tan2thetaM})
should be replaced by a plus sign, which implies that unlike the
case of neutrino oscillation in the sun, no
resonant stage exists, but an enhancement of the survival probability for
antineutrinos, as discussed in~\cite{neutri-phys}.

In analogy to the X-ray tomography, the possibility that
(anti)neutrinos could be treated as a probe to detect the inner
structure of the earth has been discussed by several authors. Two
approaches  \cite{tomography} have been suggested for different
(anti)neutrino energies.

{\em Our simplified model for an earthquake fault
zone.---}Earthquakes are caused by rapid slippage along faults and release of
energy stored in the seismic zone.
Generally, a seismic zone does not consist of one single fault. The
term " fault zone" implies that the zone is composed of several
inner faults and undergoes tremendous deformation caused by severe
shear and strain.

On May 12 of 2008, an $M_S$= 8.0 earthquake occurred in Wenchuan,
Sichuan Province, China. It was caused by the fracture of faults in the
Longmenshan fault zone. This fault zone is composed of three
parallel faults. The Wenchuan-Maoxian fault is the west most one.
According to geological studies, the Longmenshan fault zone is
stressed by the west Bayan Har Block and Sichuan Basin
\cite{yanxing-li}, and a 300 km long and 50 km wide deformation zone
was formed horizontally.


The simplified geometric structure of the fault zone, which we are
going to use in this work, is a cuboid with length $l$ = 300 km,
width $w$ = 50 km and depth $d$ = 10 km, and with an upper edge
parallel to the earth's surface.
In general, a crust extends to a depth of 30 km \cite{crust-depth},
and this cuboid is just located at its middle layer. The main shock
of the Wenchuan earthquake has a hypocenter depth of about 14 km,
and
\begin{figure}[t]
\includegraphics*[width=1\columnwidth]{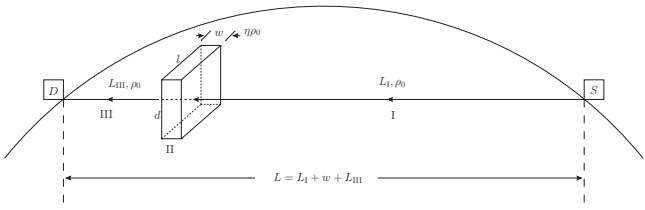}
\caption{ The configuration of the
trajectory of the (anti)neutrino beam which is produced in the
source $S$, propagates through slabs $\rm{I}$, $\rm{II}$ and
$\rm{III}$, and reaches the detector $D$. The baseline length
$L$=$L$$_{\rm I}$+$w$+$L$$_{\rm{III}}$. The geometry of the fault
zone is described by three parameters, length $l$, depth $d$ and
width $w$. The matter density of each slab is also marked in the
figure. \label{fig:geometry}}
\end{figure}
this hypocenter is inside the cuboid. A schematic of the cuboid is
shown in Fig. \ref{fig:geometry}.




The geological structure with a fault zone is not stable and
sometimes varies actively. The drastic variations within the region
of such geological structure can induce severe earthquakes. Before
an earthquake takes place,
the variation in the matter density, or more exactly, the electron
density of the fault zone is the most significant factor. The
relevant effects are categorized into four aspects:
\begin{itemize}{\setlength\itemsep{-0.5ex}
\item {\bf Strain}. Strain can change the matter density directly.
During the Wenchuan earthquake, the Longmenshan fault zone is
strained by the west and east geological blocks. The map of strain
intensity drawn by a professional research group \cite{zuan-chen}
shows that the eastern edge of the Qinghai-Xizang Plateau is close
to the region where strain intensity changes steeply. The
Longmenshan fault zone undergoes the steepest change.
The strain intensity on the west side is four times larger than that
on the east side.
\item {\bf Shear}. Shear plays an important role in the slippage of the fault.
It can either increase or decrease the matter density. From the
maximum shear strain rate map of the Bayan Har Block and Sichuan
Basin \cite{yanxing-li}, one can observe that the maximum shear
strain rate of the east of the Longmenshan fault zone is 5$\times
10^{-9}/a$, whereas on the west side, Bayan Har Block, it is 15
$\sim$ 50$\times 10^{-9}/a$.
The picture indicates that the shear contrast of both sides is 3
$\sim$ 10.
\item {\bf Anomalous electric field}. The total electron content (TEC)
anomaly, as one of the precursors of an earthquake, has been observed
from the TEC map using the GPS satellites. The authors of Ref.
\cite{TEC30-90} point out that, for strong mid-latitudinal
earthquakes, the seismo-ionospheric anomalies are manifested as a
local TEC increase or decrease in the vicinity of the forthcoming
earthquake epicenter, and this phenomenon appears several days prior
to the main shock and the TEC modification can be as large as 30
$\sim$ 90 \% of the normal value. The physical mechanism for such
ionospheric anomalies is that an anomalous electric field is
produced near the epicenter. Prior to an earthquake, intensive gas
discharges occurring at the crust in the earthquake preparation zone
lead to an anomalously strong vertical electric field \cite{TAO}.
The TEC disturbances are the vertical drift of the ionospheric
plasma under the influence of the anomalous electric field. On the
other hand, if the anomalous vertical electric field definitely
exerts influence on the earthquake preparation zone, it can also
change the distribution of free electrons in the fault zone.
In particular, the research into conductivity characteristics indicates
that a low resistance thrust fault zone in the middle layer of the
Longmenshan fault zone was observed \cite{low-resistance}.
\item {\bf Rock dehydration}. Under the influence of tremendous strain
and shear, rock would undergo a dehydration process, which  also
contributes to the density shift in the fault zone
\cite{dehydration}. }\end{itemize}

The first two aspects concern the geometrical deformation, which
contributes an accumulation of mechanical energy.
When the shear exceeded the friction strength, the faults in the
fault zone were unlocked and slipped suddenly \cite{shoubiao-zhu}.


{\em Density shift and estimation of fault zone width.---}
As antineutrinos travel inside the medium, the matter potential
$V(x)$ is linearly proportional to the electron density. For the
medium in the crust, it is reasonable to assume that $N_e$ $\approx$
$N_p$ $\approx$ $N_n$, i.e. the electron fraction number $Y_e$ = $N_
e/(N_p + N_n)$ $\approx$ 0.5, where $N_e$, $N_p$ and $N_n$
correspond to electron, proton and neutron numbers in the crust
layer, and write the electron density in the normal crust matter as
\begin{eqnarray}
n_e^0(x)={\rho_0(x)\over 2 m_N}, \label{n_e(x)}
\end{eqnarray}
where $\rho_0$ is the matter density of the normal medium and $m_N$
is the mass of nucleon. As discussed above, the electron density is
shifted in the fault zone as
\begin{eqnarray}
n_e^f=\eta n_e^0. \label{Nef}
\end{eqnarray}
As pre-earthquake activities in the fault zone are taken into
account, $\rho$ varies from $\rho_0$ into an effective matter
density $\rho_f^{\prime}$
\begin{eqnarray}
\rho_f^{\prime}=\eta \rho_0, \label{rho-f}
\end{eqnarray}
which means that in the fault zone the effective matter density
shortly before an earthquake, $\rho_f^{\prime}$, is $\eta$ times
larger than the normal matter density $\rho_0$. In fact, because the
existence of the anomalous electric field does not directly
influence the real matter density, but only contributes to the
electron density of the fault zone, the real matter density of the
fault zone  $\rho_f$ would be smaller than $\eta \rho_0$, i.e.
\begin{eqnarray}
2m_Nn_e^f=\rho_f^{\prime}=\eta \rho_0 ,\nonumber \\
2m_Nn_p^f=\rho_f<\eta\rho_0 ,\label{density-inequality}
\end{eqnarray}
where $n_p^f$ is the number density of protons in the fault zone
cuboid. The experimental observation indicates that there is a
definite relation between $\rho_f^{\prime}$ and $\rho_f$, namely as
$\rho_f^{\prime}$ increases, $\rho_f$ increases too. The
antineutrino oscillation determines $\rho_f^{\prime}$, and with the
data on $\rho_f^{\prime}$, we would obtain unambiguous information
about $\rho_f$ which causes an earthquake.



The study of geology focuses on a single fault or single fracture.
Instead, we consider a zone that consists of several faults and has
a relatively large width.
The fault zone is treated as a slab with a uniform effective matter
density $\rho_f^{\prime}$.


The width $w$ of the fault zone can also be estimated from an
empirical formula \cite{TAO}
\begin{eqnarray}
w=2\times10^{0.414M_S-1.696}\ \rm{km}, \label{radius}
\end{eqnarray}
where $M_S$ is the magnitude of the earthquake. For the Wenchuan
earthquake, $M_S$ = 8.0, the width $w$ is about 80 km from Eq.
(\ref{radius}), which is the fault zone width we use in our
numerical computations.



{\em Configuration and result for the reactor
antineutrinos.---}
The Daya Bay reactor located in Shenzhen is an example and we shall
use its corresponding data as the inputs of our numerical
computations. The fault zone cuboid in Wenchuan is about 1418 km to
Daya Bay. As is shown in the configuration figure, the $\bar \nu_e$
first traverses the normal medium with a matter density $\rho_0$ =
2.8 $g/cm^3$ and then enters the fault zone, which has a width $w$ =
50 km.

Since fault zones are not accessible for direct observation, all
information concerning details of the zones is still coarse and
sparse. Obviously, more concrete and precise information is needed
very badly to forecast disasters. Along with the geologic
study, theoretical work and new ideas are also necessary.
Certainly, abundant geological data and development in seismology
are helpful for the improvement of our fault zone picture and
further research. In this work, we set $\eta =3$ as a simple
demonstration of our scheme.

We suppose that the beam would be detected at the Qinghai-Xizang
Plateau, which is 200 km away from the fault zone. This description
of the total configuration of the trajectory is shown in Fig.
\ref{fig:geometry}.

Under the approximation of $\theta_{13} \sim 0$ the vacuum
oscillation probability $P_{\bar{\nu}_{e}\rightarrow \bar{\nu}_{e}}$
within the three-generation scenario can be reduced into the
two-generation case. When taking the matter effect into account and
considering the normal hierarchy $\bigtriangleup m^{2}_{21}\ll
\bigtriangleup m^{2}_{31} \simeq \bigtriangleup m^{2}_{32}$, the
small mass-squared difference $\bigtriangleup m^{2}_{21}$ dominates
the oscillation as the effect induced by the larger mass-squared
difference $\bigtriangleup m^{2}_{31}$ is averaged out for the long
propagation \cite{neutri-phys}. Then the evolution equation is
\begin{eqnarray}
i{{d}\over{dx}}(\hat{\psi}_{\alpha 1}, \hat{\psi}_{\alpha 2},
\hat{\psi}_{\alpha 3})^{T} = {{1}\over{2E_{\nu}}}\hat{\mathbb{M}}
(\hat{\psi}_{\alpha 1}, \hat{\psi}_{\alpha 2}, \hat{\psi}_{\alpha
3})^{T}, \label{3-evolution}
\end{eqnarray}
where $\hat{\psi}_{\alpha k}(\alpha=e,\mu,\tau; k=1,2,3)$ is the
$\nu_{\alpha}\rightarrow\nu_{k}$ effective evolution amplitude and
$\hat{\mathbb{M}}$ is given as \cite{neutri-phys}
\begin{eqnarray}
\left( \begin{array}{rrr} s^{2}_{12}\triangle
m^{2}_{21}+c^{2}_{13}A_{cc} & c_{12}s_{12}\triangle m^{2}_{21} &
-c_{13}s_{13}e^{-i\delta}A_{cc} \\ c_{12}s_{12}\triangle m^{2}_{21}
&  c^{2}_{21}\triangle m^{2}_{21} & 0
\\-c_{13}s_{13}e^{i\delta}A_{cc} & 0 & \triangle
m^{2}_{31}+s^{2}_{13}A_{cc} \end{array}  \right) , \nonumber
\end{eqnarray}
where $s_{ij}\equiv \sin\theta_{ij}$, $c_{ij}\equiv \cos\theta_{ij}$
and $A_{cc}= 2E_{\nu}V(x)$.

Under the approximation $\theta_{13}\sim 0$, the amplitude
$\hat{\psi}_{\alpha 3}$ is decoupled from the evolution equations
for $\hat{\psi}_{\alpha 1}$ and $\hat{\psi}_{\alpha 2}$, and the
effective survival probability $P^{(3)}_{\bar{\nu}_{e}\rightarrow
\bar{\nu}_{e}}$ within the three-generation scenario just reduces
into $P^{(2)}_{\bar{\nu}_e \rightarrow \bar{\nu}_{e}}$ which is the
expression in the two-generation case as
\begin{eqnarray}
P^{(3)}_{\bar{\nu}_{e} \rightarrow \bar{\nu}_{e}} = (1-
\sin^{2}\theta_{13})^{2} P^{(2)}_{\bar{\nu}_{e}\rightarrow
\bar{\nu}_{e}} + \sin^{4}\theta_{13}. \label{3->2generation}
\end{eqnarray}
It is noted that the mixing angle $\theta_{23}$ does not explicitly
show up in the survival probability of $\bar\nu_e\to\bar\nu_e$.
Since $\sin^{2}\theta_{13} < 5\times 10^{-2}$ (99.73\% CL)
\cite{sin-theta13}, with the assumption $\theta_{13}\sim 0$, the
two-generation analysis involving two parameters $\bigtriangleup
m^{2}_{21}$ and $\theta_{12}$ would be a good approximation for the
realistic case. In fact, the mixing between $\nu_{\mu}$ and
$\nu_{\tau}$ is sizable \cite{Xingzz}, therefore for the
long-distance propagation, the oscillation between $\nu_{\mu}$ and
$\nu_{\tau}$ is almost complete, i.e. $\sin^2 2\theta_{23}\approx
1$.
\begin{figure}[t]
\includegraphics*[width=1.04\columnwidth]{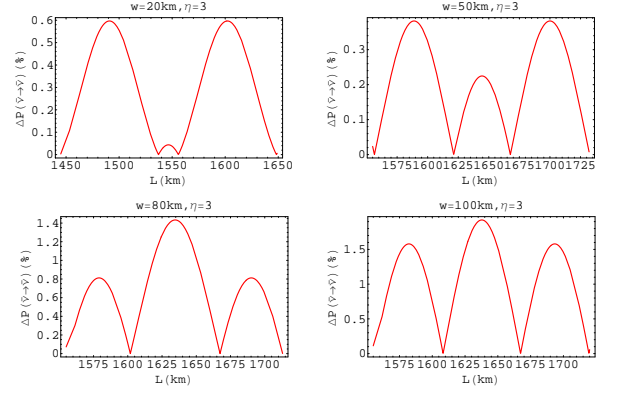}
\caption{The differences (in percent) in
survival probabilities of the reactor electron antineutrino between
cases without and with an anomalous accumulation of electrons whose
$\eta$ =3 and width $w$ = 20, 50, 80 and 100 km, respectively. \label{fig:Daya-4}}
\end{figure}
Thus for the practical mixing between $\bar{\nu}_e$ and
$\bar{\nu}_x$, $\bar{\nu}_x$ really is the superpositions of
$\bar{\nu}_{\mu}$ and $\bar{\nu}_{\tau}$. In the study of the solar
neutrino flux missing where only two generations are considered, the
updated values of the input parameters are determined as $\sin^2
2\theta_0$ = 0.86 and $\Delta m^2$ = 8$\times 10^{-5}$ $\rm{eV}^2$.
The energy of the reactor electron-antineutrino of $E_\nu$ = 3.6 MeV
\cite{neutri-phys} is another input parameter for our numerical
computation. The complete formulation involving three generations
would make the whole picture very complicated and does not change
our qualitative conclusion.

In this configuration, there are three slabs, I, II and III. Each of
them is assumed to possess a uniform matter density and can be
treated with an $R$-matrix \cite{nicolaidis}.
\begin{eqnarray}
\left(\begin{array}{c} \bar{\nu}_e^{\rm{III}} \\
\bar{\nu}_x^{\rm{III}}
\end{array} \right) &=&
R^{\rm{III}}R^{\rm{II}}R^{\rm{I}} \left(\begin{array}{c} \bar{\nu}_e^0 \\
\bar{\nu}_x^0
\end{array} \right).
\label{R-matrix}
\end{eqnarray}
The elements of the $R$-matrix are determined by the mass-squared
difference $\Delta m^2_M$, neutrino energy $E_\nu$ and
antineutrino path length in the slab. In a uniform slab, the
elements of the $R$-matrices are
\begin{eqnarray}
&R_{11}&=\exp{\left(i{\Delta m^2_M \over {2E_\nu}}x\right)} +
\sin^2{\theta_M}\left[1-\exp{\left(i{\Delta m^2_M \over
{2E_\nu}}x\right)}\right],\nonumber \\ &R_{22}&=\exp{\left(i{\Delta
m^2_M \over {2E_\nu}}x\right)} +
\cos^2{\theta_M}\left[1-\exp{\left(i{\Delta m^2_M \over
{2E_\nu}}x\right)}\right], \nonumber \\ &R_{12}&=R_{21}={1\over
2}\sin^2{2\theta_M}\left[1-\exp{\left(i{\Delta
m_M^2\over{2E_\nu}}x\right)}\right], \nonumber  \\ \label{R-element}
\end{eqnarray}
where $x$ is the antineutrino path length in the slab. we calculate
the survival probabilities without
\begin{figure}[t]
\includegraphics*[width=1.04\columnwidth]{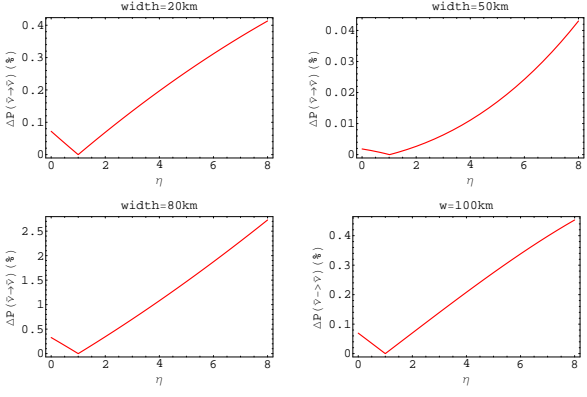}
\caption{The dependence of the
differences (in percent) in survival probabilities between cases
without and with an anomalous accumulation of electrons on the
$\eta$ value for reactor antineutrinos whose width $w$ = 20, 50, 80
and 100 km, respectively. \label{fig:width-eta}}
\end{figure}
and with the anomalous accumulation of electrons at the fault zone.
In Fig. \ref{fig:Daya-4} we present the differences in survival
probabilities of the reactor electron antineutrino between cases
without and with an anomalous accumulation of electrons for $\eta$
=3 and different widths. For oscillations in medium with and without
the anomalous electron accumulation, the maximal difference in
$P_{\bar{\nu}_e\rightarrow \bar{\nu}_e}$ is up to 2\% when the width $w$
= 100 km.

Among all of the fault zone parameters, the width and the $\eta$ value
play important roles in antineutrino oscillations. For reactor
electron antineutrinos, the dependence of the differences of
survival probabilities on $\eta$ is plotted in Fig.
\ref{fig:width-eta} with antineutrino energy $E_\nu$ = 3.6 MeV for
different width values. Fig. \ref{fig:width-eta} indicates that for
$\eta < 1$, the effect is tiny, and for $\eta$ is larger than one,
the predicted probability difference undergoes a relatively larger
change with the increase in $\eta$.




{\em Neutrino factory.---}
Neutrino beams from a neutrino factory \cite{neutrino-factory} can
also be employed in the neutrino oscillation tomography.
In our configuration, we choose an electron neutrino beam with
energy $E_{\nu}$ = 7 GeV.
Our calculation indicates that the difference in survival
probabilities corresponding to the cases with and without the
anomalous electron accumulation is not obvious. Indeed, an
observable difference of about 0.01\% may be detected if the
detector is 6000 km away from the source. However, such a length of
baseline is much beyond the maximal distance in the crust layer and
such a small difference may not be detected with the
\begin{table}[h]
\begin{center}
\begin{tabular}{|c||c|c|c|}\hline
Parameter & \multicolumn{3}{c|}{ Reactor }\\  \hline Slab & I
& II & III \\ \hline $L$ (km) & 1418 & $w$ & 200\\ \hline$ \rho$ $
(g/\rm{cm}^3)$ & 2.8 & $2.8 \times3 $ & 2.8\\ \hline $\Delta
m^2_{\rm M}$ $(10^{-5}\rm{eV}^2)$ & $8.02$ & $8.09$ & $8.02$\\
\hline $\tan 2\theta_M$ & 2.42 & 2.30 &
2.42\\ \hline $L^M_{osc}$ $(\rm{km})$ & 111.18 & 110.36 & 111.18\\
\hline\hline Parameter & \multicolumn{3}{c|}{ Neutrino Factory }\\
\hline Slab & I & II & III
\\ \hline $L$ (km) & 1418 & $w$ & 200\\ \hline $ \rho$ $
(g/\rm{cm}^3)$ & 2.8 & $2.8 \times3 $ & 2.8\\ \hline $\Delta
m^2_{\rm M}$ $(10^{-3}\rm{eV}^2)$ & $1.46$ & $4.42$ & $1.46$\\
\hline $\tan 2\theta_M(10^{-2})$ & $-5.10$ & $-1.68$ & $-5.10$\\
\hline $L^M_{osc}$ $(10^3\rm{km})$ & $11.92$ & $3.92$ & $11.92$\\
\hline
\end{tabular}
\begin{quote}
\caption{The parameters of neutrino
oscillation in matter for beams produced in a reactor and in a neutrino
factory. For reactor electron antineutrinos $E_\nu$ = 3.6 MeV and
for neutrino factory electron neutrinos $E_\nu$ = 7 GeV.} \label{tab:osc-parameters}
\end{quote}
\end{center}
\end{table}
accuracy of the present neutrino facilities in our configuration.
The reason is obvious that the accelerator neutrinos have much
larger energy than the reactor antineutrinos, thus the oscillation
length $L_{osc}^M$ for neutrinos from a neutrino factory is much
longer than that for the reactor ones as shown in Table
\ref{tab:osc-parameters}.


{\em Result from perturbation method.---}
In a low matter density medium, $V \ll {\Delta m^2}/{2E_{\nu}}$,
Ioannisian and Smirnov \cite{perturbation} developed a perturbation
method where $\epsilon = 2VE_{\nu}/\Delta m^2$ is taken as the
expansion parameter for calculating the oscillation probability. For
the maximal matter density $\eta \rho _0$ in our case, one can
determine
\begin{eqnarray}
\frac{2E_{\nu}V}{\Delta m^2}\approx0.029\ll1, \label{2EV/m}
\end{eqnarray}
for $E_{\nu}$ = 3.6 MeV and $\eta$ = 3. With the perturbation
method, the exact adiabatic phase shift in matter is obtained as


\begin{eqnarray}
\phi^m_{a\rightarrow
b}=\frac{1}{2E_{\nu}}\int_a^bdx&\biggl[&\left(\Delta \cos
{2\theta_0}\pm\sqrt{2}E_{\nu}G_F\frac{\rho
\left(x\right)}{m_N}\right)^2
\nonumber\\
&& +\left(\Delta \sin{2\theta_0}\right)^2\biggr]^{1\over 2}
\label{fiab}
\end{eqnarray}
and the plus and minus signs correspond to antineutrinos and
neutrinos, respectively. By the formula for flavor-to-flavor
transition, we obtain the survival probability of the electron
antineutrino originating from the initial spot $x_0$ and ending at
the final spot
\begin{figure}[t]
\includegraphics*[width=1\columnwidth]{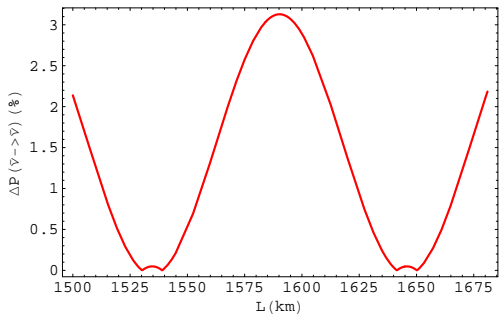}
\caption{The difference (in percent) of
survival probabilities between cases without and with an anomalous
accumulation of electrons evaluated in the perturbation theory
Eq.~(\ref{P-pertur}). \label{firstorder}}
\end{figure}
$x_f$. Up to the leading order of $V(x)$, we have
\begin{eqnarray}
P_{\bar{\nu}_e\rightarrow \bar{\nu}_e}=&1&-\sin^2{2\theta
_0}\biggl[\sin^2{\frac{1}{2}\phi ^m_{x_0\rightarrow x_f}}
-\frac{\sqrt{2}}{4}\cos{2\theta_0}\nonumber\\
&\times& \int_{x_0}^{x_f}dx \frac{G_F\rho \left(x\right)}{m_N}
\left(\sin{\phi_{x_0\rightarrow x}^{m}+\sin{\phi _{x\rightarrow
x_f}^{m}}}\right) \biggr]. \nonumber\\ \label{P-pertur}
\end{eqnarray}
In this scheme, we calculate the survival probability for reactor
electron antineutrinos. The result is shown in Fig.
\ref{firstorder}. The difference in survival probabilities for the
cases with and without the anomalous electron accumulation can be as
large as 3$\%$, which is larger than the results obtained with the
$R$-matrix method.

{\em Discussion.---}
We note that there is a sharp controversy about the $\eta$ value
among geologists. Some think that this value should be rather small,
whereas others advocate that, before a
severe earthquake, the $\eta$ might reach a sizable value. Taking the
advice of the more conservative geologists, we adopt a
relatively small value in this work. Indeed, we only investigate the
physics and provide a possible way to forecast the earthquake in
terms of neutrino medium effects in this work, but we will leave the
detailed study of the $\eta$ value to the geologists.

In fact, the other neutrino sources, such as the cosmic muon,
would contaminate our detection circumstance. However, we may set
some shielding and veto facilities to reduce the  uncertainty caused
by other neutrino sources. The detailed technical difficulties would
be left for our experimental colleagues.



Foreseeing earthquakes in a certain region within ten years, two
years and one or two months corresponds to long-, medium- and
short-term earthquake forecast, while an impending earthquake
forecast should be made a few days before the outbreak. For the
long-term and medium-term forecast, there is quite a long duration
for the detector to collect sufficient data even though the flux is
suppressed by the small solid angle spanned by the detector. For
short-term and impending forecasting, the electron density changes
drastically as the earthquake is coming, and the overall effect
results in a large shift in the oscillation probability in a short
time interval, so that observing clear signals might be feasible if
the flux from the source is sufficiently strong. As mentioned previously,
the geometrical deformation caused by strain and shear would last
for a long period, say, hundreds of years in some cases and the
deformation would eventually reach its maximum before the outbreak.
However, the strong anomalous electric field appears only several
days, e.g. about a week before the earthquake, so that the week
before the outbreak is crucial for detectors to collect data and
reach a conclusion about the upcoming disaster.

Now let us make a rough numerical evaluation of the event rates
based on the Daya Bay Monte-Carlo simulation. In the Daya Bay case,
the detector is set at 1 km away from the neutrino source, and the
detection rate is about a few hundred per day ($\sim$400/day, as
estimated). Supposing a shortened baseline in our case is 200km, the
suppression factor for the flux would be $1/(200)^2\sim 3\times
10^{-5}$. Considering an exposure time of 100 days and assuming that the
detection efficiency is 100 times larger than that employed in
Daya Bay experiment, the expected events would be at least $\mathcal
{O}$(100), which is suitable for a medium-term forecast. For
example, in a period of one year, one can compare the events
observed in the first half of the year with that in the second half. If
these data show an explicit difference in survival probabilities
during the two periods, they would imply a geological deformation
that may result in an earthquake in the forthcoming period. The
closer the earthquake is, the larger the difference.

Indeed, in our present work, we consider idealized beam
characteristics while ignoring many details. The uncertainty caused
by other neutrino sources such as other reactors and geoneutrinos,
is also ignored, which may be one of the main challenges in the
implementation of this scheme. Thus we reach our conclusion that
with the present reactor facilities and detection techniques, it is
still a difficult task to make an earthquake forecast using such a
scheme, though it seems to be possible from a theoretical point of
view while ignoring some uncertainties. However, there is hope
that with the great improvement in detection and innovation of
facilities, the sensitivity and efficiency would be greatly improved
and the size of the detector may be much larger with sufficient fund
support, and with the development of the geology and seismology, the
pictures of the fault zone will be clearer and its parameters
more accurate, thus a medium-term earthquake forecast would be
feasible.

A scheme might remedy the shortcoming of remarkably losing neutrino
flux due to the small solid angle. Namely, one may use a low energy
beta beam \cite{beta-beam} as a source. Such a neutrino beam could
be directly oriented to the fault zone. Without the $10^{-5}$
suppression factor, the requirement for the size and sensitivity of
the detectors would be greatly alleviated. However, in order to
obtain intense and collimated neutrino beams from the traditional
source for the beta beam, e.g., $ ^{6}_{2}\rm{He}^{++} {\rightarrow}
^{6}_{3}\rm{Li}^{+++} + \rm{e}^{-} + \bar{\nu}_{e}$
\cite{beta-beam}, the high intensity of the $^{6}_{2}\rm{He}^{++}$
source and a large Lorentz boost factor are necessary. We suggest
alternative schemes. One is that in the process
$e^{-}+\rm{p}\rightarrow \rm{n}+\nu_{e}$ we can employ the very
collimated electron beam to bombard the protons at rest. Another is
that we can use the lowly accelerated pion beam for
$\pi^{\pm}\rightarrow e^{\pm}+\bar{\nu}_{e}(\nu_{e})$. Even though
in the second proposal, the decay rate is suppressed by the helicity
rule, the branching ratio is still of the order of $10^{-4}$
\cite{Data}, so with a large number of pions, it is still plausible.
Moreover, a large Lorentz boost factor $|{\bf p}|/m_e$ which
determines the spreading solid angle, is easy to obtain and the
energy for the pion accelerator needs to be just at
$\mathcal{O}(\rm{GeV})$. Thus the size of such facilities can be
small and the corresponding technique would be relatively simple.

\begin{acknowledgments}
We sincerely thank Prof. B.L. Young who introduced
the applications of neutrino beams in various fields, especially
the neutrino tomography, to us. We have greatly benefited from very
enlightening discussions on the subject with Prof. Chung Ngoc Leung.
In fact, this work was motivated by the discussion when he visited
the Nankai campus. This work is supported by the National Natural
Science Foundation of China (NNSFC) and the Special Grant of the
Education Ministry of China for the Ph.D programs.
\end{acknowledgments}

\end{document}